\documentclass[prb,twocolumn,showpacs,floatfix,amsmath,amssymb,superscriptaddress]{revtex4-2}
\usepackage{amsfonts}
\usepackage{stmaryrd}
\usepackage{bbm}
\usepackage{mathrsfs}
\usepackage{tipa}
\usepackage{amssymb}
\usepackage{txfonts}
\usepackage{graphicx}
\usepackage{dcolumn}
\usepackage{epstopdf}
\usepackage[colorlinks,linkcolor=black,urlcolor=black,citecolor=black]{hyperref}
\usepackage{multirow}
\usepackage{subfigure}
\usepackage{url}
\usepackage{upgreek}
\usepackage[utf8]{inputenc}
\usepackage[english]{babel}
\usepackage[locale = US]{siunitx}
\usepackage{lipsum}

\begin{document}

\title{High-order nonlinear terahertz probing of the two-band superconductor MgB$_2$: Third- and fifth-order harmonic generation}

\author{C. Reinhoffer}
\affiliation{Institute of Physics II, University of Cologne, 50937 Cologne, Germany}

\author{P. Pilch}
\author{A. Reinold}
\affiliation{Department of Physics, TU Dortmund University, 44227 Dortmund, Germany}

\author{P.~Derendorf}
\affiliation{Institut für Theoretische Physik III, Ruhr-Universität Bochum, 44801 Bochum, Germany}

\author{S. Kovalev}
\author{J.-C. Deinert}
\author{I.~Ilyakov}
\author{A.~Ponomaryov}
\author{Min~Chen}
\affiliation{Institute of Radiation Physics, Helmholtz-Zentrum Dresden-Rossendorf, 01328 Dresden, Germany}

\author{Tie-Quan Xu}
\affiliation{Applied Superconductivity Center and State Key Laboratory for Mesoscopic Physics, School of Physics, Peking University, Beijing 100871, China}

\author{Yue~Wang}
\affiliation{Applied Superconductivity Center and State Key Laboratory for Mesoscopic Physics, School of Physics, Peking University, Beijing 100871, China}
\affiliation{Peking University Yangtze Delta Institute of Optoelectronics, Nantong, Jiangsu 226010, China}

\author{Zi-Zhao~Gan}
\affiliation{Applied Superconductivity Center and State Key Laboratory for Mesoscopic Physics, School of Physics, Peking University, Beijing 100871, China}
\affiliation{Peking University Yangtze Delta Institute of Optoelectronics, Nantong, Jiangsu 226010, China}

\author{De-Sheng~Wu}
\affiliation{Beijing National Laboratory for Condensed Matter Physics, Institute of Physics, Chinese Academy of Sciences, Beijing, 100190 China}

\author{Jian-Lin~Luo}
\affiliation{Beijing National Laboratory for Condensed Matter Physics, Institute of Physics, Chinese Academy of Sciences, Beijing, 100190 China}
\affiliation{Songshan Lake Materials Laboratory, Dongguan, Guangdong, 523808 China}

\author{S.~Germanskiy}
\author{E.~A.~Mashkovich}
\author{P.~H.~M.~van~Loosdrecht}
\affiliation{Institute of Physics II, University of Cologne, 50937 Cologne, Germany}

\author{I.~M.~Eremin}
\affiliation{Institut für Theoretische Physik III, Ruhr-Universität Bochum, 44801 Bochum, Germany}

\author{Zhe Wang}
\email{corresponding author: zhe.wang@tu-dortmund.de}
\affiliation{Department of Physics, TU Dortmund University, 44227 Dortmund, Germany}

\date{\today}

\begin{abstract}
We report on high-order harmonic generation in the two-band superconductor MgB$_2$ driven by intense terahertz electromagnetic pulses. Third- and fifth-order harmonics are resolved in time-domain and investigated as a function of temperature and in applied magnetic fields crossing the superconducting phase boundary.
The high-order harmonics in the superconducting phase reflects nonequilibrium dynamics of the superconducting order parameter in MgB$_2$, which is probed via nonlinear coupling to the terahertz field.
The observed temperature and field dependence of the nonlinear response allows to establish the superconducting phase diagram.
\end{abstract}

\maketitle

\section{Introduction}

High-harmonic spectroscopy has been demonstrated to be very powerful in revealing novel properties of matter, especially by providing time-resolved dynamical characteristics for states far from thermodynamic equilibrium \cite{CorkumKrausz,GhimireReis19,Ghimire2011,Schubert2014,Luu2015,Vampa2015,You2017,
Liu2017,Soavi2018,Hafez2018,Kovalev19,Cheng20,Bai2021,Lv2021,Dantas21,Kemper13,lara_CDF1,
phonon_retard_higgs,Rubio17,
Silva2018,Takayoshi2019,Schwarz20,impurity_MgB2,
haenel2021timeresolved,muller2021,Shao2022, Matsunaga_NbN_Science,Rajasekaran18,
JGWang_NPhoto19,Chu20,uchida2021highorder,NbN_FHG_22}.
From these high-order nonlinear characteristics one can realize dynamical Bloch oscillations 
\cite{Ghimire2011,Schubert2014,Luu2015,Soavi2018}, reconstruct electronic band structures \cite{Vampa2015,You2017,Rubio17}, and identify topological and relativistic effects \cite{Liu2017,Kovalev19,Cheng20,Bai2021,Lv2021,Dantas21,Germanskiy22}.
For a more complex system with strong electron-electron interactions, 
high-harmonic generation opens up new opportunities to disclose fingerprints for exotic states of matter, such as amplitude (Higgs) modes \cite{PekkerVarma15,ShimanoTsuji20}, stripe phases in cuprates \cite{Rajasekaran18}, light- or current-driven symmetry breaking phases \cite{JGWang_NPhoto19,Jigang2020,Nakamura20}, and light-induced phase transitions in strongly correlated systems \cite{Silva2018,Shao2022}.
Moreover, ellipticity dependence of high-harmonic generation allows to probe magnetic dichroism \cite{Kfir15,Fan15} and molecular chirality on a sub-picosecond time scale \cite{Cireasa15}.

Recent developments on accelerator-based intense and high-repetition-rate terahertz source have enabled high-harmonic spectroscopic studies in the few terahertz-frequency range (1~THz $\sim$~4.1~meV) with femtosecond time resolution  \cite{Kovalev17,Hafez2018,Kovalev19,Chu20}. In this work we use the terahertz (THz) high-harmonic spectroscopy to investigate characteristic nonlinear response of the representative two-band superconductor MgB$_2$.

With a layered crystalline structure, MgB$_2$ consists of boron atoms forming a primitive honeycomb lattice and magnesium atoms located above the center of the boron hexagons in-between the boron planes \citep{Akimitsu01}.
The Fermi surface in MgB$_2$ is characterized by a two-dimensional $\sigma$-band (bonding $p_{x,y}$ bands) and a three-dimensional $\pi$-band (bonding and antibonding $p_z$ bands) \cite{MazinPRL01a}.
Below a superconducting transition at about 40~K, two superconducting energy gaps open simultaneously in the $\sigma$- and $\pi$-bands with $2\Delta_\sigma=13$~--~14 and $2\Delta_\pi=3$~--~5~meV, respectively, at the lowest temperatures \cite{MgB2_twogap_STM,Iavarone_STM_2002,MgB2_twogap_ARPES}.
With a stronger intraband scattering the $\pi$ band is in the dirty limit \cite{Koshelev03,Eskildsen02,vdMPRB06},
whereas the $\sigma$ band with a smaller scattering than $2\Delta_\sigma$ is intermediately clean (see e.g. \cite{vdMPRB06,MgB2Kovalev}).
As a type-II superconductor MgB$_2$ maintains a Meissner phase up to a lower critical field $H_{c1}\sim 0.1$~T (see e.g.~\cite{Tan15}),
while the upper critical field $H_{c2}$ exhibits evident anisotropy with respect to magnetic field orientation (see e.g.~\cite{Eltsev02,Welp03}). Moreover, the resistive superconducting transitions are not sharp in applied magnetic fields, which may reflect metastability of the vortex system around $H_{c2}$ \cite{Eltsev02,Welp03}.
In resistivity measurements for $H \parallel c$ the onset of superconductivity determines $H^{\text{onset}}_{c2}=7.0$~--~7.5~T, while the zero resistivity occurs at $H^{\text{zero}}_{c2}=3.0$~--~3.5~T for the lowest temperatures (see e.g. \cite{Eltsev02,Welp03} and Appendix Fig.~\ref{Fig_S1_MR}). Ordered vortex-lattice phases were resolved well below $H^{\text{zero}}_{c2}$ \cite{Das12}. For $H \parallel ab$ the upper critical field extends up to above 20~T \cite{Eltsev02,Welp03}.

With a relatively high $T_c$ and a representative two-band nature, MgB$_2$ as a model system has been vividly investigated by various pump-probe spectroscopic techniques (see e.g. \cite{Xu2003, Demsar2003, Baldini2017, Giorgianni2019, Novko2020, MgB2Kovalev}). For example, nonequilibrium dynamics of optically excited carriers was studied by optical reflection probe \cite{Xu2003} or THz probe \cite{Demsar2003} with sub-ps time resolution, in which the important role of electron-phonon couplings has been revealed. 
A white-light probe of the optically excited transient states disclosed fingerprint for interband scattering between the $\sigma$- and $\pi$-bands \cite{Baldini2017}, and found a particular $E_{2g}$ phonon mode which interacts strongly with the $\sigma$-bands \cite{Baldini2017,Novko2020}. More recently, THz field driven nonequilibrium states of MgB$_2$ was investigated by THz probe \cite{Giorgianni2019} or by harmonic generation \cite{MgB2Kovalev}, which provided possible evidence for Leggett mode or Higgs mode, respectively. Since these experiments were performed for different frequencies and waveforms of pump pulses, a more systematic study remains necessary to clarify the involved nonequilibrium dynamics \cite{Fiore2022}.

Following our previous work on temperature dependence of THz third-harmonic generation in MgB$_2$ in zero magnetic field \cite{MgB2Kovalev}, here we carry out studies of THz third-harmonic generation in applied external magnetic fields and also observe fifth-order harmonic generation as a function of temperature. These investigations provide a more systematic characterization of the THz nonlinear response of the superconducting state in MgB$_2$.

\section{Experimental details}

High-quality single-crystalline MgB$_2$ thin films with the \textit{c}-axis epitaxy and a thickness of 10~nm were grown on $5\times 5$ mm MgO (111) substrate by using a hybrid physical-chemical vapor deposition method, and characterized by x-ray diffraction and charge transport measurements \cite{WangYue13}. 
Magnetic field dependence of \textit{dc} resistivity was measured with a current of 100~µA at various temperatures for applied fields parallel to the \textit{c}-axis, using a physical properties measurement system (PPMS-9T, Quantum Design).

The THz harmonic generation measurements were performed at the TELBE user facility in the Hemholtz Zentrum Dresden-Rossendorf.
Intense narrow-band THz pulses were generated based on a linear electron accelerator which was operated at 50~kHz and synchronized with an external femtosecond laser system \cite{Kovalev17}.
The THz radiation was detected by electro-optic sampling in a ZnTe crystal using the synchronized laser system.
Bandpass filters with central frequencies of a bandwidth of 20~\% were applied to produce narrow-band THz radiation of desired frequencies.
The bandwidth here is defined as a full-width at the half-maximum of transmission, which corresponds to +-10~\% of the central frequency. Well beyond the central frequency the transmission is suppressed by about $30$~dB.
The experiment was aligned in a transmission geometry. The emitted harmonic radiation was detected after the sample in the direction of the normally incident THz beam.
For the excitation and detection of high-harmonic generation, $f$-, $3f$- and $5f$-bandpass filters with central frequencies of $f=0.3$, $3f=1$, $f=0.4$ and $5f=2.1$~THz were used.
Magnetic field dependent measurements were carried out in a cryomagnet (Oxford Instruments) with the magnetic field applied along the \textit{c}-axis of the MgB$_2$ thin films, the same orientation as for the \textit{dc} resistance measurements.

Since the MgB$_2$ samples degrade in air, we sealed a fresh piece of sample under vacuum before the user experiment. During the experiment the sample was installed in an evacuated cryostat. The spectroscopic data presented here was acquired for a continuous 84 hours experiment. The transport data were obtained on a different piece of sample.

\section{Experimental results and discussion}

\subsection{Experimental results}

\begin{figure}[t]
\includegraphics[width=8cm]{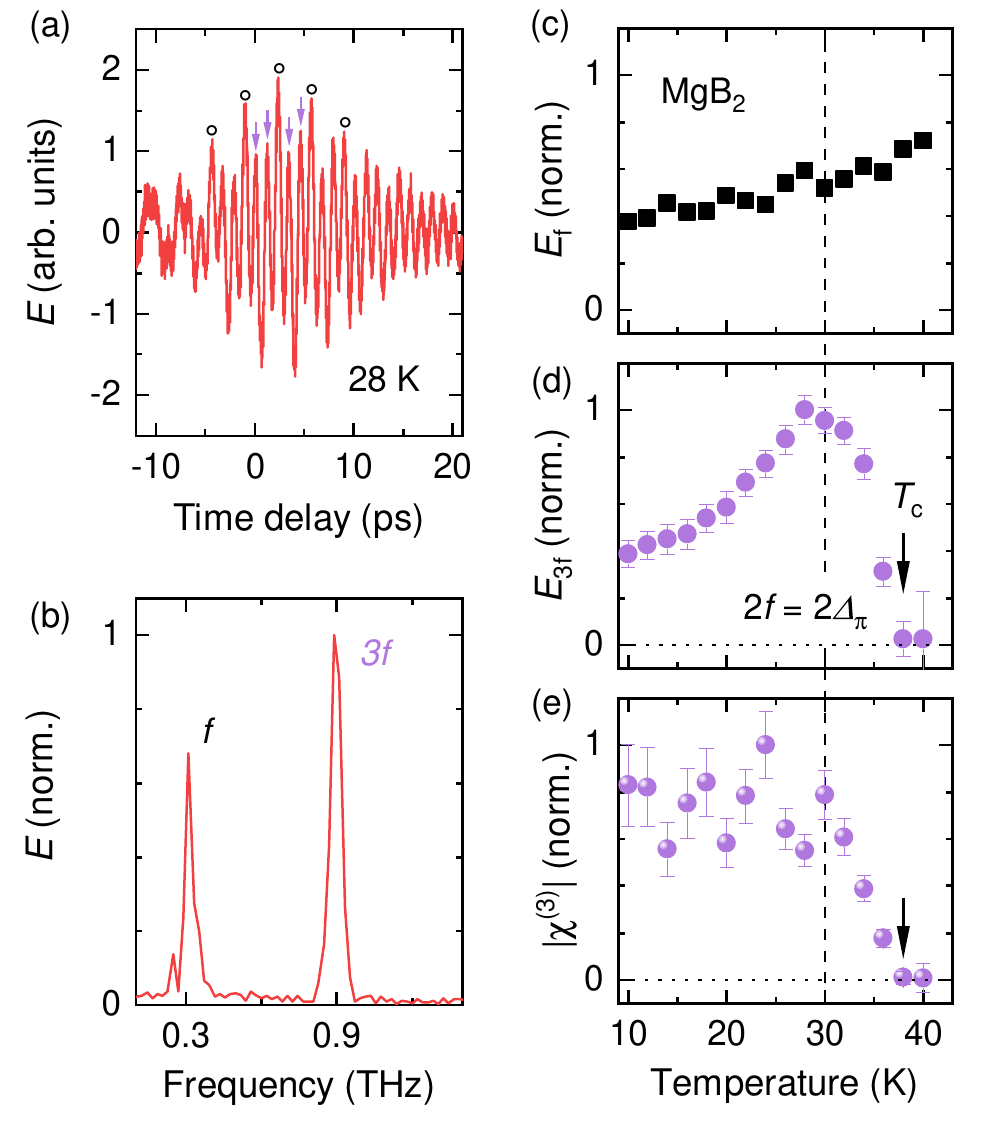}
\caption{\label{Fig_300GHz_Tdep}
(a) Emitted THz electric field from MgB$_2$ driven by a $f=0.3$~THz pulse at 28~K and in zero magnetic field. The data is recorded through a $3f$-bandpass filter with a central frequency of 1~THz. The arrows indicate the $3f$ signal.
(b) Fourier spectrum of the time-domain signal in (a) exhibits third-harmonic radiation at $3f = 0.9$~THz.
(c) $f$- and (d) $3f$-components of the emitted electric field as a function of temperature.
(e) $|\chi^{(3)}|\equiv E_{3f}/E^3_f$ versus temperature. 
The dashed line indicates the temperature with the resonance condition $2f=2\Delta_{\pi}$ fulfilled.} 
\end{figure}

We start with reproducing the previously reported third-harmonic generation \cite{MgB2Kovalev}, since we are now using a different THz source i.e. based on an electron accelerator and measuring a different piece of sample. 
Under the drive of a 0.3~THz excitation pulse with peak electric field of 13~kV/cm, the electric field of emitted radiation from MgB$_2$ was recorded in time-domain, which for 28~K below $T_c$ is presented in Fig.~\ref{Fig_300GHz_Tdep}(a). 
In addition to the major peaks (as marked by the circles) that correspond to the driving pulse, one can clearly see two additional peaks [see arrows in Fig.~\ref{Fig_300GHz_Tdep}(a)] between every two neighbouring major peaks.
This is a direct observation of third-harmonic radiation from the superconducting state of MgB$_2$ on sub-picosecond time scale.
By performing Fourier transformation of the time-domain signal, the obtained spectrum exhibits two sharp peaks at $f=0.3$~THz and $3f=0.9$~THz, respectively [Fig.~\ref{Fig_300GHz_Tdep}(b)].
We note that the time-domain signal is measured through a $3f$-bandpass filter, which suppresses the $f$ components substantially, leading to the apparent larger amplitude at $3f$ in Fig.~\ref{Fig_300GHz_Tdep}(b).  

The amplitude of the $f$ and $3f$ components ($E_f$ and $E_{3f}$) as derived from Fig.~\ref{Fig_300GHz_Tdep}(b) is presented in Fig.~\ref{Fig_300GHz_Tdep}(c) and Fig.~\ref{Fig_300GHz_Tdep}(d), respectively, as a function of temperature.
While the monotonic decrease of $E_f$ with reduced temperature reflects increased screening in the superconducting phase \cite{CarrHomes01,MgB2_THz_TDS,Pronin01,BBJin05,vdMPRB06,Ortolani08,MgB2Kovalev}, the third-harmonic radiation $E_{3f}$ exhibits a broad maximum around 30~K.
Above $T_c$ the third-harmonic yield is essentially zero, whereas dramatically enhanced below the superconducting transition. Below 30~K, $E_{3f}$ starts to drop again with decreasing temperature.
Such a decrease might have simply resulted from the enhanced reflection of the driving field well below $T_c$, since the third-order response should be more sensitive to the screening of the driving field.
However, this cannot fully explain the observed maximum in the temperature dependence curve. 
We evaluate temperature dependence of a third-order susceptibility $|\chi^{(3)}|$ via normalized $E_{3f}/E^3_f$, as presented in Fig.~\ref{Fig_300GHz_Tdep}(e).
With decreasing temperature from $T_c$, $|\chi^{(3)}|$ increases monotonically until 30~K, while at lower temperatures $|\chi^{(3)}|$ levels off. At 30~K, the resonance condition $2f=2\Delta_\pi$ is fulfilled \cite{Iavarone_STM_2002}, i.e. twice the pump frequency equals the superconducting energy gap in the $\pi$ band.
Hence, the observed temperature dependence of the third-harmonic generation reflects the nonlinear response of the superconductivity, rather than just being a consequence of the enhanced screening of the driving field.

\begin{figure*}[t]
\includegraphics[width=12cm]{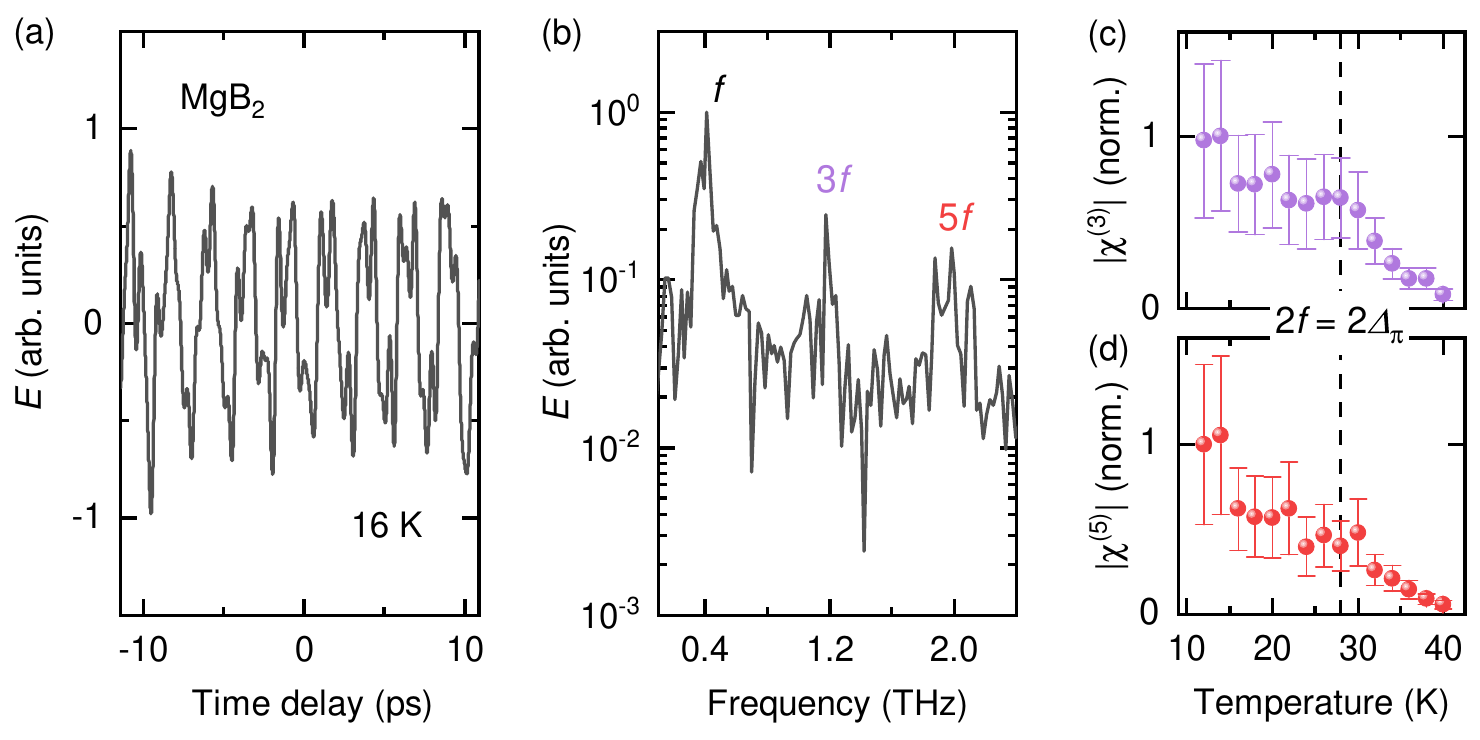}
\caption{\label{fig_FHG}
(a) Terahertz electric field emitted from MgB$_2$
driven by a $f=0.4$~THz pulse at 16~K in zero field.
(b) The corresponding spectrum exhibits fundamental $f$, third $3f$, and fifth $5f$ harmonic generation.
A $5f$-bandpass filter with a central frequency of 2.1~THz was applied for the detection.
Temperature dependence of (c) third- and (d) fifth-order nonlinear susceptibility, $|\chi^{(3)}|\equiv E_{3f}/E^3_f$ and $|\chi^{(5)}|\equiv E_{5f}/E^5_f$, respectively.
Dashed line indicates the temperature corresponding to the resonance condition $2f=2\Delta_{\pi}$.}
\end{figure*}

High-harmonic radiation even up to the fifth order is observed from the superconducting state of MgB$_2$
under a drive of 0.4~THz pulse with a peak electric field of 60~kV/cm, which can be seen directly from the time-resolved electric field of the radiation [Fig.~\ref{fig_FHG}(a)].
This is more evident in its Fourier spectrum [Fig.~\ref{fig_FHG}(b)], which displays clearly the fundamental, third-, and fifth-harmonic components.
The absence of even-order harmonics is dictated by the existence of an inversion symmetry in the crystal structure of MgB$_2$ with a space group of \textit{P6/mmm} \cite{Akimitsu01,Boyd2020}.
Here we have employed a $5f$-bandpass filter to enable a simultaneous detection of the different components, from which we can estimate the temperature dependence of the nonlinear susceptibilities, as presented in Fig.~\ref{fig_FHG}(c) and Fig.~\ref{fig_FHG}(d) for normalized $|\chi^{(3)}|\equiv E_{3f}/E^3_f$ and $|\chi^{(5)}|\equiv E_{5f}/E^5_f$, respectively.
Finite values of the nonlinear susceptibilities appear below $T_c$, characterizing nonlinear response of the superconducting state.
With decreasing temperature an initial rapid increase of $|\chi^{(3)}|$ and $|\chi^{(5)}|$ is followed by a gentle one below about 28~K or rather level-off behaviour towards the lowest temperature.
As indicated by the dashed line in Fig.~\ref{fig_FHG}(c)(d),
this temperature corresponds to the same resonance condition, $2f=2\Delta_\pi$, for $f=0.4$~THz and the $\pi$-band gap $2\Delta_\pi$ \cite{Iavarone_STM_2002}. 
We note that neither the $3f$ nor the $5f$ signals observed here are just leakage of the pump pulse through the bandpass filters, otherwise they should appear also well above the superconducting transition with even higher intensity because of reduced screening of the pump pulse.  

\begin{figure}[t]
\includegraphics[width=8.7cm]{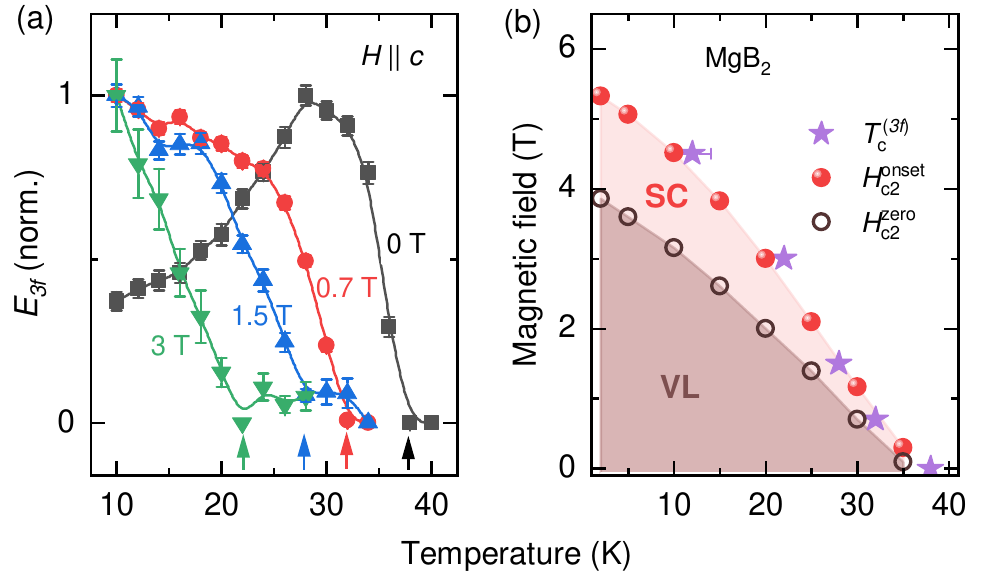}
\caption{\label{Fig_THG_Fdep}
(a) THG amplitude $E_{3f}$ as a function of temperature in various applied magnetic fields $H \parallel c$. The arrows indicate the THG onset temperature $T_c^{(3f)}$. 
(b) Magnetic field-temperature phase diagram of MgB$_2$.
$H^{\text{zero}}_{c2}$ corresponds to zero resistance (see Appendix Fig.~\ref{Fig_S1_MR}), below which ordered vortex-lattice (VL) phase is formed \cite{Das12}.
$H^{\text{onset}}_{c2}$ are the fields where the resistance starts to drop (see Appendix Fig.~\ref{Fig_S1_MR}).
The field-dependent $T_c^{(3f)}(H)$, in agreement with $H^{\text{onset}}_{c2}(T)$, determines the superconducting (SC) phase boundary.
}
\end{figure}

As presented above, the high-order nonlinear probe is very sensitive to the superconductivity of MgB$_2$. Hence we can utilize the third-harmonic generation to study the nonlinear response of the superconducting state in an applied magnetic field.
Figure~\ref{Fig_THG_Fdep}(a) displays the electric-field amplitude of emitted third-harmonic radiation versus temperature for $f=0.3$~THz in various magnetic fields along the \textit{c}-axis ($H \parallel c$).
With increasing field from 0~T, the onset of the third-harmonic generation shifts continuously towards lower temperature [see arrows in Fig.~\ref{Fig_THG_Fdep}(a)]. Moreover, the resonance-peak feature observed in zero field is replaced in finite fields by a monotonic increase of the third-harmonic generation with decreasing temperature.
Since the high-harmonic radiation reflects time-dependent spatial average of the nonlinear current over a mm beam spot size, the absence of a resonance feature may result from spatial inhomogeneity and metastability of the vortex system, besides the suppression of superconductivity in fields. 

The onset temperature of the third-harmonic generation as a function of the applied magnetic field $T^{(3f)}_c(H)$ is shown in Fig.~\ref{Fig_THG_Fdep}(b), and compared with the temperature dependence of characteristic fields $H^{\text{zero}}_{c2}(T)$ and $H^{\text{onset}}_{c2}(T)$, which corresponds to the occurrence of zero resistance and the field where the resistance starts to drop, respectively (see Appendix Fig.~\ref{Fig_S1_MR}).
One can see that $T^{(3f)}_c(H)$ matches very well with $H^{\text{onset}}_{c2}(T)$ rather than $H^{\text{zero}}_{c2}(T)$. Thus, the THz nonlinear response provides a sensitive probe of the superconducting stiffness in the applied magnetic fields.
While below $H^{\text{zero}}_{c2}$ ordered vortex lattice is formed \cite{Eskildsen02}, the finite resistance between $H^{\text{zero}}_{c2}$ and $H^{\text{onset}}_{c2}$ probably results from vortex motion under excitation of electric current \cite{Eltsev02,Welp03}. 

\subsection{Discussion}

Third-order harmonic generation in MgB$_2$ thin films has been reported  previously for the GHz frequencies (see e.g. \cite{Andreone2002,Booth2003,GALLITTO2003,Oates2008,Anlage2012}).
These low-energy responses are not only sensitive to intrinsic properties of the superconductor which can be ascribed to backflow of thermally excited unpaired quasiparticles in the presence of the superflow \cite{Dahm2004,Sauls95}, but also to extrinsic properties e.g. weak links or vortex motion \cite{Samoilova1995,Andreone2002,Booth2003,GALLITTO2003,Oates2008,Anlage2012}. In particular, the backflow of the unpaired quasiparticle is responsible for the observed GHz harmonic response mainly around $T_c$. In contrast to these previous studies, our present work focuses on the nonlinear response in the THz frequencies, whose photon energy is comparable to the superconducting gaps of this system at low temperatures and about three orders of magnitude higher than the microwaves. Moreover, the THz pulses with intense electric field drive the system out of equilibrium, thus the observed nonlinearity reflects the high-energy nonequilibrium properties of the superconductor.

On the origin of the THz high-harmonic generation in conventional superconductors there is an ongoing discussion at present. In fact in a BCS-type single-band superconductor it was found that the activation of the Higgs mode is not the only process which can be probe through the high-harmonic generation. In particular, the Bogoliubov quasiparticles give further contribution to the high-harmonic generation at $2\Delta$, as it coincides with the onset of
the particle-hole continuum corresponding to the energy needed to break the Cooper pairs. 
It was shown in Ref.~\cite{lara_CDF1} that in the clean BCS single-band superconductor the charge-density-fluctuation contribution, associated with Bogoliubov quasipartciles, to the third-harmonic generation current is three orders of magnitude larger than that from the Higgs mode. Subsequent studies analyzed how impurities affect this ratio, and the current consensus is that amplitude (Higgs) fluctuations contribution may dominate the THG signal at sufficient disorder \cite{Jujo2018,impurity_MgB2,impurity_Higgs,haenel2021timeresolved,Seibold21}.

Very recently the analysis for the third-order harmonic generation has been expanded to a two-band model \cite{Fiore2022}. The main conclusion of this study was that the third-order harmonic signal at a lower gap will be still dominated by the contribution of charge density fluctuations even if one accounts for the disorder, included in a semi-phenomenological way. To see whether this is really the case one would be required to use a quasiclassical Green's functions formalism, which goes well beyond the present mostly experimental work. Instead, in what follows we outline the framework of calculating the fifth-harmonic generation current in a clean case using a pseudospin formalism.

The Heisenberg equations of motion for the pseudospins $\vec{\sigma}_k$ are given by
\begin{equation}
\partial_t\vec{\sigma}_k(t)=i[\hat{H},\vec{\sigma}_k]=\vec{b}_k\times \vec{\sigma}_k(t)
\end{equation}
where the Hamiltonian $\hat{H}=\sum_k \vec{b}_k \vec{\sigma}_k$ with the pseudomagnetic field $\vec{b}_k$. We calculate the time evolution of the pseudospins up to the fourth order in the vector potential $A(t)$ by expanding the $z$ component of $\vec{b}_k$ in the nonlinear response regime
\begin{equation}
\epsilon_{k-eA(t)}+\epsilon_{k+eA(t)}=2\epsilon_{k}+\epsilon^{A^2}_{k}+\frac{1}{12}\epsilon^{A^4}_{k}+O(A^6)
\end{equation}
with $\epsilon_{k}$ representing the dispersion relation,
$\epsilon^{A^2}_{k}=\sum_{i,j=\{x,y\}}e^2A_i(t)A_j(t)\partial_{k_i}\partial_{k_j}\epsilon_{k}$,
and 
$\epsilon^{A^4}_{k}=\sum_{i,j,l,m=\{x,y\}}e^4A_i(t)A_j(t)A_l(t)A_m(t)\partial_{k_i}\partial_{k_j}\partial_{k_l}\partial_{k_m}\epsilon_{k}$.

Restricting the analysis to a BCS-type single-band superconductor in the clean limit (see Appendix), we find that either due to Higgs amplitude fluctuations or Cooper-pair breaking (charge density fluctuations), third-order harmonic yield exhibits a resonance enhancement for a resonance condition $2f=2\Delta$, whereas enhanced fifth-order harmonic generation occurs at two resonance conditions $4f=2\Delta$ and $2f=2\Delta$ (see Appendix Fig.~\ref{Fig_S2_FHG}). 
In the clean limit, the fifth-harmonic generation is primarily due to the Cooper-pair breaking, similar to previous theoretical analysis for third-harmonic generation (see e.g. \cite{lara_CDF1,Schwarz20,Giorgianni2019}). 
In contrast, in the dirty limit the third-harmonic generation could be dominated by Higgs amplitude fluctuations (see e.g.
\cite{phonon_retard_higgs,impurity_MgB2,haenel2021timeresolved}).
As for the amplitude fluctuations, the resonance condition $2f=2\Delta$ indicates a resonant excitation of a Higgs mode through two photons, which is coupled to another one or three photons, leading to the third- or fifth-harmonic generation, respectively.  
Corresponding to the resonance condition $4f = 2\Delta$, a Higgs mode can also be resonantly excited through four photons, resulting in the fifth-harmonic generation.

An important experimental finding here is that a resonance feature in the third- and fifth-harmonic generation [Fig.~\ref{fig_FHG}(c)(d)] is observed only for $2f=2\Delta_\pi$ corresponding to the $\pi$-band gap in the dirty limit, although for the clean-limit $\sigma$-band gap the resonance conditions $2f=2\Delta_{\sigma}$ and $4f=2\Delta_{\sigma}$ should be experimentally accessible.
These results do not support an interpretation of the observed high-harmonic generation in MgB$_2$ as being predominantly due to pair-breaking of the clean-limit band, but rather suggests that Higgs amplitude fluctuations in the dirty-limit band mainly lead to the high-harmonic signals.
Nonetheless, this argument assumes negligible inter-band couplings, and is essentially based on an independent band picture. A rigorous analysis of a two-band model by taking into account interband couplings and sufficient disorder is still required to elucidate the different contributions, not only for the third-order harmonic generation, but also for high-order harmonic generation as observed here.

\section{Conclusion}
By performing terahertz high-order harmonic spectroscopy of the superconducting state in the two-band superconductor MgB$_2$, we revealed characteristic nonlinear response of the superconductivity.  
As a function of temperature and applied magnetic field, we investigated third- and fifth-order harmonic generation in MgB$_2$ driven by intense terahertz field, and established 
its superconducting phase diagram.
Resonance enhancement of the third- and the fifth-harmonic signals in zero field was observed only for the dirty-limit $\pi$-band, i.e. $2f=2\Delta_\pi$, below the superconducting transition temperature, but not for $2f=2\Delta_\sigma$ or $4f=2\Delta_\sigma$ of the clean-limit $\sigma$-band.
While in a single-band picture this suggests a dominant contribution of the Higgs amplitude fluctuations to the high-harmonic generation in the dirty limit, the analysis of a more realistic two-band model for MgB$_2$ by taking into account the interband couplings and disorder is still necessary for a more reliable interpretation of our experimental results.

\begin{acknowledgments}
We thank L. Benfatto, T. Dahm, M. Eskildsen, J. Fiore, and J. Kierfeld for very helpful discussions.
The work in Cologne was partially supported by the DFG via Project No. 277146847 — Collaborative Research Center 1238: Control and Dynamics of Quantum Materials (Subproject No. B05).
Parts of this research were carried out at ELBE at the Helmholtz-Zentrum Dresden - Rossendorf e.V., a member of the Helmholtz Association. We acknowledge partial support by MERCUR (Mercator Research Center Ruhr) via Project No. Ko-2021-0027.
\end{acknowledgments}

\section{Appendix}

\subsection{Pseudospin analysis}
To understand the signatures of the superconducting state and its collective modes in the fifth-harmonic generation (FHG) we follow the standard approach of using the Anderson pseudospin outlined in \cite{Derendorf22},
which is defined as:
\begin{equation}
    \Vec{\sigma}_k = \frac{1}{2}\Psi_k^\dagger\Vec{\tau}\Psi_k
\end{equation}
Here, $\Vec{\tau}$ denotes the Pauli-matrices vector and $\Psi_k=\begin{pmatrix}\hat{c}_{k\uparrow} & \hat{c}_{-k\downarrow}\end{pmatrix}^T$ is the Nambu-Gorkov spinor. The BCS Hamiltonian can be written in the compact form
\begin{equation}
    \hat{H}_{BCS}=\sum_k \Vec{b}_k\Vec{\sigma}_k
\end{equation}
with the pseudomagnetic-field $\Vec{b}_k = 2\begin{pmatrix}-\Delta^{'} & \Delta^{''} & \epsilon_k\end{pmatrix}^T$, where $\Delta=\Delta^{'}+i\Delta^{''}$. The gap equation can also be expressed via pseudospins $\Delta=V\sum_k\left(\left<\hat{\sigma}^{x}_{k}\right>+i\left<\hat{\sigma}^{y}_{k}\right>\right)$, where $\hat{\sigma}^{x,y}_k$ denote the component of the pseudospin vector. In the following we restrict our analysis here to a single gap in the clean limit.   

To calculate the third- and fifth-order nonlinear currents, we first compute the gap-oscillation under irradiation whose vector potential is given by $A(t)=A_0\sin{\left(\Omega t\right)}$ with $\Omega=2\pi f$ the angular frequency. We include the coupling of the electronic system to the electromagnetic field via a Peierls-substitution, which affects the $z$-component of the pseudomagnetic-field
\begin{equation}
    \Vec{b}_k = \begin{pmatrix}-2\Delta^{'} & 2\Delta^{''} & \epsilon_{k+eA}+\epsilon_{k-eA}\end{pmatrix}^T.
\end{equation}
The time-evolution of the pseudospins obeys the Heisenberg equations of motion 
\begin{equation}\label{Eq:timePseudo}
    \partial_t\Vec{\sigma}_k(t)=i\left[\hat{H},\Vec{\sigma}_k\right] = \Vec{b}_k\times\Vec{\sigma}_k(t)
\end{equation}
Following the standard procedure we linearize the equations by expressing $\Vec{\sigma}_k(t)=\left<\Vec{\sigma}_k(0)\right>+\delta\Vec{\sigma}_k(t)$ and $\Delta(t)=\Delta+\delta\Delta(t)$ and decouple the resulting system of differential equation by transforming Eq.\,(\ref{Eq:timePseudo}) to Laplace-space. We further set $\Delta$ to be real at $t=0$ which allows to  express the equilibrium's values for the pseudospin components

\begin{equation}
\begin{split}
    \left<\hat{\sigma}_k^x(0)\right>&=\frac{\Delta}{2E_k}\tanh{\left(\frac{E_k}{2k_BT}\right)}\\
    \left<\hat{\sigma}_k^y(0)\right>&=0\\
    \left<\hat{\sigma}_k^z(0)\right>&=\frac{-\epsilon_k}{2E_k}\tanh{\left(\frac{E_k}{2k_BT}\right)},
\end{split}    
\end{equation}
where $E_k=\sqrt{\epsilon_k^2+\Delta^2}$ is the Bogoliubov energy dispersion. To compute $\delta\Delta^{'}(s)$ and $\delta\Delta^{''}(s)$ we expand $\epsilon_{k+eA}+\epsilon_{k-eA}$ up to the fourth-order in $A$. Evaluating the obtained expression we obtain both $\delta\Delta^{'}$ and $\delta\Delta^{''}$:
\begin{equation}\label{Eq:Delta(s)}
\begin{split}
    \delta\Delta^{'}(s)&=\frac{1}{2}\left(\alpha_1\Delta e^2 A^2(s)+\frac{1}{12}\Tilde{\alpha}_1\Delta e^4 A^4(s)\right)\left(1-\frac{1}{\lambda(4\Delta^2+s^2)F(s, T)}\right)\\
    \delta\Delta^{''}(s)&=\underbrace{-\frac{\alpha_0}{s}\Delta e^2A^2(s)}_{=\delta\Delta^{''}_{A^2}(s)}\underbrace{-\frac{\Tilde{\alpha}_0}{12s}\Delta e^4A^4(s)}_{=\delta\Delta^{''}_{A^4}(s)}
\end{split}
\end{equation}
where we separate $\delta\Delta^{'}(s)=\delta\Delta^{'}_{A^2}(s)+\delta\Delta^{'}_{A^4}(s)$ into two parts. 
The factors $\alpha_{0}=\frac{-\epsilon_F a^2}{2}$, $\Tilde{\alpha}_{0}=\frac{\epsilon_F a^4}{2}$, $\alpha_{1}=\frac{-a^2}{2}$, $\Tilde{\alpha}_{1}=\frac{a^4}{2}$ originate from the nearest-neighbour hopping term on a square lattice with a lattice constant denoted by $a$. For simplicity we set $a=1$.
The function $F(s, T)$ is given by:
\begin{equation}
    F(s, T) = \int\mathrm{d}\epsilon\frac{\tanh{\left(\frac{E}{2k_BT}\right)}}{2E(4E^2+s^2)}
\end{equation}
For simplicity we have assumed that $\Vec{A}(t)$ is parallel to the lattice vector.

Finally the current can be expressed via the $z$-component of the pseudospin as $\vec{j}(t)= \vec{v}_{k-eA}\left(\left<\hat{\sigma}_k^z(t)\right>+\frac{1}{2}\right)$. We expand $\vec{v}_{k-eA}$ up to the third-order in $A$, transform $\vec{j}$ into Fourier space, and by plugging the expression for the gap oscillation we arrive at the following six terms, contributing to the FHG.

The Higgs contribution $\propto \delta\Delta^{'}(s)$ towards the fifth-order nonlinear current is given by
\begin{equation}
\begin{split}
    \vec{j}^{(5)H}_\parallel(5\Omega)&=-ie^2A_0\sum_k\frac{\Delta_0(\alpha_0+\alpha_1\xi_k)\tanh{\left(\frac{E}{2k_BT}\right)}}{E_k(4E_k^2-16\Omega^2)}2\xi_k\delta\Delta'_{A^4}(s=4i\Omega)\\
    &+\frac{ie^4A_0^3}{24}\sum_k\frac{\Delta_0(\Tilde{\alpha}_0+\Tilde{\alpha}_1\xi_k)
\tanh{\left(\frac{E}{2k_BT}\right)}}{E_k(4E_k^2-4\Omega^2)}2\xi_k\delta\Delta'_{A^2}(s=2i\Omega).
\end{split}
\end{equation}
where $\xi_k = \epsilon_k-\epsilon_F$ with $\epsilon_F$ the Fermi energy.
Thus we obtain two components to the order of $A^2$ or $A^4$, respectively,
\begin{equation}
    \begin{split}
        j_{A^2}^H(5\Omega)&=i\frac{\alpha_1\Tilde{\alpha}_1\Delta^2e^6A_0^3A^2(2i\Omega)}{24}\left(1-\frac{H(2i\Omega, T)}{2}-\frac{1}{2H(2i\Omega, T)}\right),\\
        j_{A^4}^H(5\Omega)&=-i\frac{\alpha_1\Tilde{\alpha}_1\Delta^2e^6A_0A^4(4i\Omega)}{12}\left(1-\frac{H(4i\Omega, T)}{2}-\frac{1}{2H(4i\Omega, T)}\right),
    \end{split}
\end{equation}
where we define $H(s, T)=\lambda(4\Delta^2+s^2)F(s, T)$.

\begin{figure}[t]
\includegraphics[width=8.5cm]{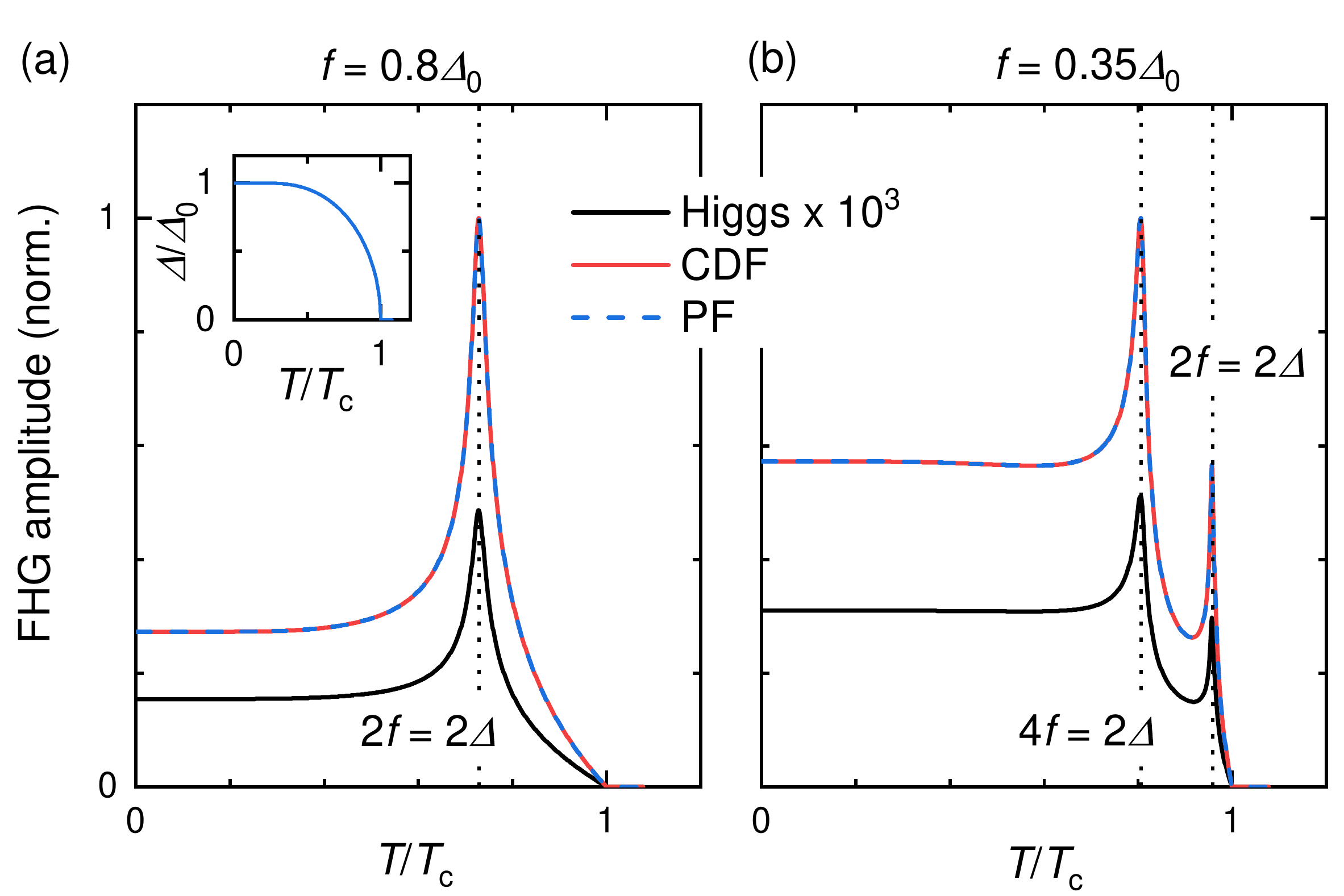}
\caption{\label{Fig_S2_FHG}
Amplitude of fifth-harmonic generation due to Higgs mode, charge-density fluctuations (CDF), and phase fluctuations (PF) for (a) $f = 0.8\Delta_0$ and (b) $f = 0.35\Delta_0$ in the clean limit. Inset: BCS-type temperature dependence of the superconducting gap.
}
\end{figure}

The contribution to FHG due to the phase fluctuations originates from the sum containing $\propto\delta\Delta{''}(s)$, i.e.
\begin{equation}
\begin{split}
    \vec{j}^{(5)P}_\parallel(5\Omega)&=ie^2A_0\sum_k\frac{\Delta_0(\alpha_0+\alpha_1\xi_k)
\tanh{\left(\frac{E}{2k_BT}\right)}}{E_k(4E_k^2-16\Omega^2)}4i\Omega\delta\Delta''_{A^4}(s=4i\Omega)\\
    &-\frac{ie^4A_0^3}{24}\sum_k\frac{\Delta_0(\Tilde{\alpha}_0+\Tilde{\alpha}_1\xi_k)
\tanh{\left(\frac{E}{2k_BT}\right)}}{E_k(4E_k^2-4\Omega^2)}2i\Omega\delta\Delta''_{A^2}(s=2i\Omega)
\end{split}
\end{equation}
Hence
\begin{equation}
    \begin{split}
        j_{A^2}^P(5\Omega)&=-i\frac{1}{12}\alpha_0\Tilde{\alpha}_0\Delta^2e^6A_0^3A^2(2i\Omega)\lambda F(2i\Omega, T),\\
        j_{A^4}^P(5\Omega)&=i\frac{1}{6}\alpha_0\Tilde{\alpha}_0\Delta^2e^6A_0A^4(4i\Omega)\lambda F(4i\Omega, T).\\
    \end{split}
\end{equation}
Finally, the BCS contribution due to Cooper-pair breaking (often dubbed charge density fluctuations) is
\begin{equation}
\begin{split}
    \vec{j}^{(5)\mathrm{CDF}}_\parallel(5\Omega)&=ie^2A_0\sum_k\frac{\Delta_0^2(\alpha_0+\alpha_1\xi_k)\tanh{\left(\frac{E}{2k_BT}\right)}}{E_k(4E_k^2-16\Omega^2)}\frac{1}{12}\xi_k^{A^4}(s=4i\Omega)\\
    &-\frac{ie^4A_0^3}{24}\sum_k\frac{\Delta_0^2(\Tilde{\alpha}_0+\Tilde{\alpha}_1\xi_k)\tanh{\left(\frac{E}{2k_BT}\right)}}{E_k(4E_k^2-4\Omega^2)}\xi_k^{A^2}(s=2i\Omega)
\end{split}
\end{equation}
Two separate components are given by
\begin{equation}
    \begin{split}
j_{A^2}^{\mathrm{CDF}}(5\Omega)&=-i\frac{1}{12}\Delta^2e^6A_0^3A^2(2i\Omega)\alpha_0\Tilde{\alpha}_0\lambda F(2i\Omega, T)\\
&-i\frac{1}{48}\Delta^2e^6A_0^3A^2(2i\Omega)\alpha_1\Tilde{\alpha}_1(1-H(2i\Omega, T))\\
        j_{A^4}^{\mathrm{CDF}}(5\Omega)&=i\frac{1}{6}\Delta^2e^6A_0A^4(4i\Omega)\alpha_0\Tilde{\alpha}_0\lambda F(4i\Omega, T)\\
        &+i\frac{1}{24}\Delta^2e^6A_0A^4(4i\Omega)\alpha_1\Tilde{\alpha}_1(1-H(4i\Omega, T))
    \end{split}
\end{equation}

One observes that some of the terms are almost identical to that appearing in the third-harmonic generation (THG) \cite{Schwarz20}. The new terms are those proportional to $A^4(4i\Omega)$.

Since the experimentally measured THz electric field is proportional to the time derivative of the current density, we compute the Fourier amplitude $\mathcal{F}\left(\partial j(t)/\partial t\right)$, which are just the expressions above with a leading factor of $5\Omega$. For the numerical computation we use  $\lambda=0.25$, $T_D=\SI{500}{K}$, $t=10\cdot\Delta_0$, $\epsilon_F=\SI{-200}{meV}$ and a broadening factor $\gamma=0.025\Delta_0$ where $2\Delta_0=3.2$~meV denotes the superconducting gap at zero temperature, and assume a BCS-type temperature dependence of superconducting gap (see the inset of Fig.~\ref{Fig_S2_FHG}).

The computed FHG amplitude of various contributions (Higgs amplitude mode, charge density fluctuations, and phase fluctuations) is presented in Fig.~\ref{Fig_S2_FHG}(a)(b) for pump-pulse frequencies $f= 0.8\Delta_0$ and $0.35\Delta_0$, respectively. The charge density fluctuations and phase fluctuations provide the same contribution to the FHG. 
For $f= 0.8\Delta_0$ which is greater than $\Delta_0/2$, we observe only one resonant peak in the FHG located at the temperature where  $2\Delta(T)=2f$. This is similar to the resonance condition for THG, meaning that this is the THG process carrying over into the FHG.
In contrast, for frequencies below $\Delta_0/2$ such as $f= 0.35\Delta_0$, we find two distinct peaks at the temperatures $T_1$ and $T_2$, where $2\Delta(T_1)=2f$ and $2\Delta(T_2)=4f$. 
The additional peak at $T_2$ arises as the resonant excitation of a Higgs mode through four photons.
In addition, we see that the BCS contribution dominates the FHG in the clean limit, as expected also for THG (see e.g. \cite{Schwarz20,lara_CDF1}).

\subsection{dc resistance in applied magnetic field}

\begin{figure}[t]
\includegraphics[width=7cm]{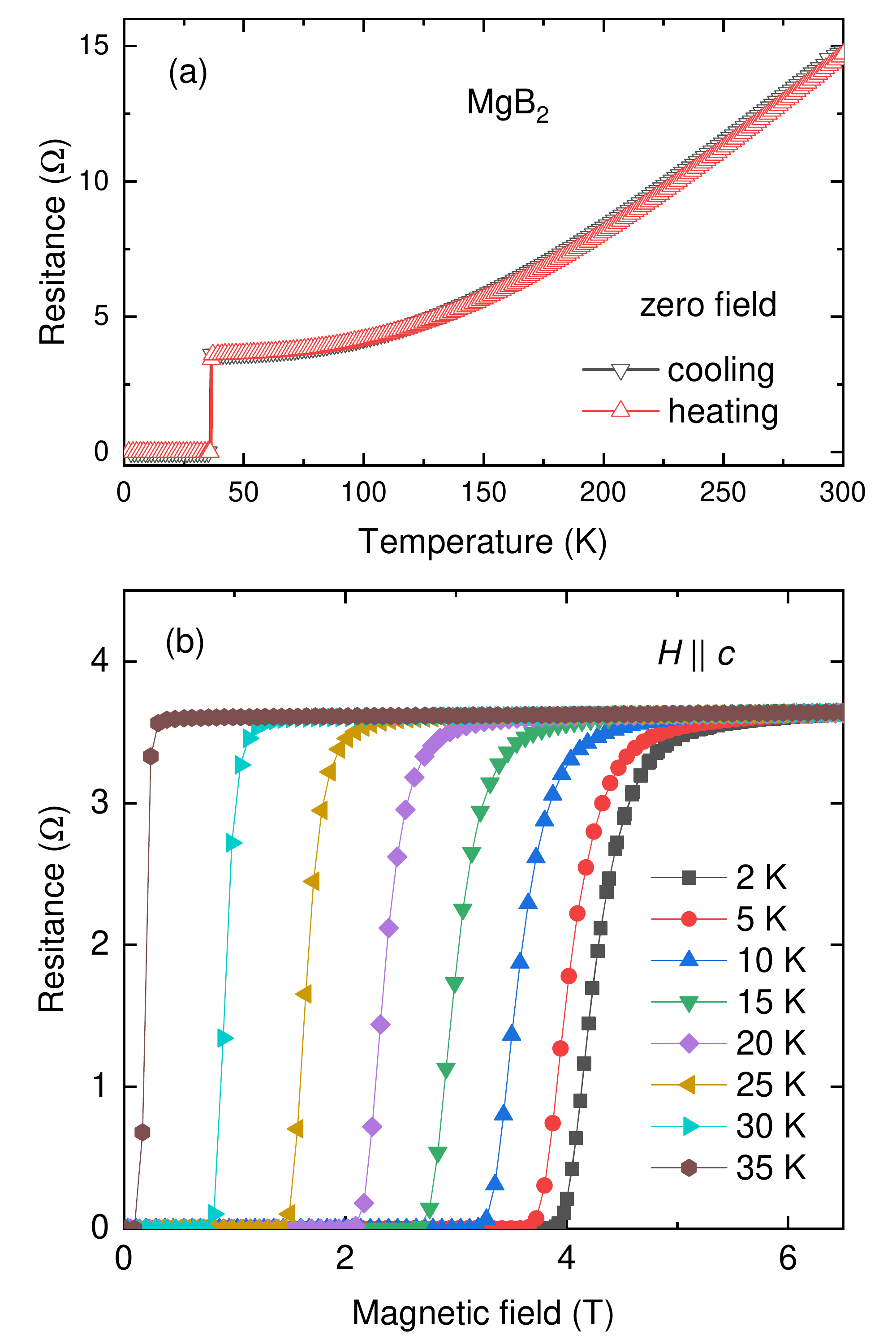}
\caption{\label{Fig_S1_MR}
(a) temperature-dependent \textit{dc}-resistance of a MgB$_2$ sample in zero field and (b)
\textit{dc} resistance measured upon field sweeping at different temperatures. 
}
\end{figure}

Figure~\ref{Fig_S1_MR} presents (a) temperature-dependent \textit{dc}-resistance of a MgB$_2$ sample in zero field and (b) field-dependent \textit{dc}-resistance measurements for various temperatures with the magnetic field applied along the crystallographic $c$-axis, which is the same orientation as applied in the high-harmonic generation experiment.

\bibliographystyle{apsrev4-2}
\bibliography{MgB2_bib}

\newpage


\end{document}